\documentclass[aps,prl,twocolumn,superscriptaddress,floatfix]{revtex4-2}

\usepackage{lineno}
\usepackage{pdftexcmds}
\usepackage{graphicx}  
\usepackage{dcolumn}   
\usepackage{bm}        
\usepackage{amssymb}   
\usepackage{amsmath}
\usepackage{blkarray, multirow, graphicx, diagbox, color, xcolor, colortbl}
\usepackage{bbm, bbold}
\usepackage{ifthen}
\usepackage[colorlinks, linkcolor = blue, citecolor = blue, filecolor = black, urlcolor = blue]{hyperref}

\usepackage{xkeyval}
\usepackage{moreverb}
\usepackage{rotating}
\usepackage{slashbox}
\usepackage{xspace}
\usepackage{capt-of}
\usepackage[caption=false]{subfig}
\newcommand\subfigref[1]{\protect\subref{#1}}

\usepackage{glossaries}
\glsdisablehyper 
\usepackage{hyphenat}
\usepackage{nicefrac}
\usepackage[]{units}
\usepackage{physics}
\usepackage{braket}
\usepackage[inline]{enumitem}
\usepackage{tabto}
\usepackage{listings}
\usepackage{xstring}
\def\ReplaceStr#1{%
	\IfSubStr{#1}{p}{%
		\StrSubstitute{#1}{p}{.}}{#1}}
\def\ReReplaceStr#1{%
	\IfSubStr{#1}{.}{%
		\StrSubstitute{#1}{.}{p}}{#1}}
\usepackage{gnuplottex}
\usepackage{tikz}
\usepackage{calc}
\usepackage{pgffor}
\usepackage{pgfplots}
\pgfplotsset{compat=1.13}
\usepackage{pgfplotstable}
\usepgfplotslibrary{groupplots}
\usepgfplotslibrary{fillbetween}

\tikzstyle{n} = [draw,shape=ellipse,minimum size=1.5em,inner sep=0pt,fill=white!20, minimum width=2.5em]
\tikzstyle{Init} = [n,color=green,fill=green!20,text=black]
\tikzstyle{Fin} = [n,color=red,fill=red!20,text=black]
\tikzstyle{Ghost} = [minimum size=1.5em,inner sep=0pt,color=white,text=black]
\tikzstyle{Multiple} = [draw,shape=rect,minimum size=2em,inner sep=0pt]

\tikzstyle{ghostA} = [text=red!70,thick, minimum size=2*(5pt-\pgflinewidth), inner sep=0pt, outer sep=0pt]
\tikzstyle{ghostB} = [text=blue!70,thick, minimum size=2*(5pt-\pgflinewidth), inner sep=0pt, outer sep=0pt]
\tikzstyle{siteA} = [draw=red!70,circle,thick, minimum size=2*(5pt-\pgflinewidth), inner sep=0pt, outer sep=0pt]
\tikzstyle{siteB} = [draw=blue!70,circle,thick, minimum size=2*(5pt-\pgflinewidth), inner sep=0pt, outer sep=0pt]
\tikzstyle{operatorA} = [cross out, draw=red!70, thick, minimum size=2*(5pt-\pgflinewidth), inner sep=0pt, outer sep=0pt]
\tikzstyle{operatorB} = [cross out, draw=blue!70, thick, minimum size=2*(5pt-\pgflinewidth), inner sep=0pt, outer sep=0pt]

\definecolor{colorA}{rgb} {0.58,0,0.8275}
\definecolor{colorB}{rgb} {0.11,0.663,0.51}
\definecolor{colorC}{rgb} {0.3373,0.7059,0.9137}
\definecolor{colorD}{rgb} {0.902,0.6235,0}
\definecolor{colorE}{rgb} {0.9451,0.902,0.3255}
\definecolor{colorIPphase}{rgb} {0.4, 0.7, 0.4}
\definecolor{colorIPcharge}{rgb} {0.7, 0.1, 0.7}

\def\gpmarkers{{"+","x","star","square","square*","o","*","triangle","triangle*"}}
\def\gpcolors{{"colorA","colorB","colorC","colorD","colorE"}}

\def\MIMmarkers{{"o","x","square","star","square*","o","*","triangle","triangle*"}}
\def\MIMcolors{{"black","magenta","cyan","colorD","colorE"}}

\usetikzlibrary
{
	calc,
	decorations,
	fadings,
	plotmarks,
	patterns,
	positioning,
	petri,
	arrows,
	decorations.markings,
	backgrounds,
	fit,
	graphs,
	shapes.geometric,
	decorations.pathmorphing,
	shapes.misc,
	shapes,
	tikzmark,
	pgfplots.colorbrewer,
	fpu
}

\pgfplotsset{
        cycle from colormap manual style/.style={
            x=3cm,y=10pt,ytick=\empty,
            stack plots=y,
            every axis plot/.style={line width=2pt},
        },
}

\tikzset
{
	style matrix_highlight_green/.style=
	{
		set fill color=green!20,
		set border color=green!50,
		fill opacity=0.5
	},
	style matrix_highlight_red/.style=
	{
		set fill color=red!20,
		set border color=red!50,
		fill opacity=0.5
	},
	style matrix_highlight_blue/.style=
	{
		set fill color=cyan!20,
		set border color=blue!40
	},
	hor/.style=
	{
		above left offset={-0.15,0.5},
		below right offset={0.15,-0.15},
		#1
	},
	hor2/.style=
	{
		above left offset={-0.15,0.4},
		below right offset={0.15,-0.15},
		#1
	},
	hor3/.style=
	{
		above left offset={-0.15,0.25},
		below right offset={0.15,0.0},
		#1
	},
	ver/.style=
	{
		above left offset={-0.1,0.5},
		below right offset={0.15,-0.15},
		#1
	}
}

\tikzset{->-/.style={decoration={
			markings,
			mark=at position .5 with {\arrow{>}}},postaction={decorate}}}

\tikzstyle{orientedsnake} = [
decorate, 
decoration={snake},
->
]  
\tikzstyle{orientedshortarrow} = [
decoration={markings,
	mark=at position .33 with {\arrow{>}}},
postaction={decorate}
]  
\tikzstyle{orientedlongarrow} = [
decoration={markings,
	mark=at position .67 with {\arrow{>}}},
postaction={decorate}
]
\tikzset{dbl/.style={double,
		double equal sign distance,
		-implies,
		shorten >=10pt,
		shorten <=10pt}}
\tikzset{
	between/.style args={#1 and #2}{
		at = ($(#1)!0.5!(#2)$)
	}
}
\pgfmathdeclarefunction{quadFct}{3}%
{%
	\pgfmathparse{#1*x*x+#2*x+#3}%
}
\pgfmathdeclarefunction{linearFct}{2}%
{%
	\pgfmathparse{#1*x+#2}%
}
\pgfmathdeclarefunction{logFct}{2}%
{%
	\pgfmathparse{#1*log10(x)+#2}%
}
\pgfmathdeclarefunction{expFct}{2}%
{%
	\pgfmathparse{#1*exp(-x/#2)}%
}
\pgfmathdeclarefunction{ccFct}{3}
{%
	\pgfmathparse{(#1)/6.0*ln((#2)*sin(deg((pi*x)/(#2))))+(#3)}%
}%

\newcommand{\nodagger}[0]{\phantom{\dagger}}

\pgfmathdeclarefunction{peierlspotential}{6}%
{
	\pgfmathparse
	{
		#2 / (#3 * #5) 
		* sin(deg(#1 * #3 - #3 * #5 * (x)))
		* exp(-( ( #1 - #5 * ((x) - #4) ) )^2.0 / ( #6^2.0 ) )
	}%
}
\newboolean{buildCurrentBlockInline}
\setboolean{buildCurrentBlockInline}{false}
\newboolean{rebuildCurrentBlock}
\setboolean{rebuildCurrentBlock}{false}

\newboolean{rebuildData}
\setboolean{rebuildData}{false}

\newboolean{buildtikzpics}
\setboolean{buildtikzpics}{true}

\newboolean{useShellEscape}
\setboolean{useShellEscape}{true}

\newif\ifrebuildtikz
\newif\ifChangeMode
\ChangeModetrue
\ChangeModefalse
\usetikzlibrary{external}
\ifthenelse{\boolean{useShellEscape}}
{
	\rebuildtikztrue
}
{
	\rebuildtikzfalse
	\tikzset{external/export=false}
}

\usepackage{ulem}
\usepackage{cleveref}
\Crefname{appendix}{Appendix}{Appendices}
\Crefname{equation}{Equation}{Equations}
\Crefname{figure}{Figure}{Figures}
\Crefname{section}{Section}{Sections}
\Crefname{paragraph}{Paragraph}{Paragraphs}
\Crefname{tabular}{Tabular}{Tabulars}
\crefname{appendix}{App.}{Apps.}
\crefname{equation}{Eq.}{Eqs.}
\crefname{figure}{Fig.}{Figs.}
\crefname{section}{Sec.}{Secs.}
\crefname{paragraph}{Par.}{Pars.}
\crefname{tabular}{Tab.}{Tabs.}

\ExplSyntaxOn
\DeclareExpandableDocumentCommand \eval { m } { \fp_eval:n { #1 } }
\ExplSyntaxOff

\newcommand{\printpgfnumberwitherror}[2]%
{%
	\pgfmathfloatparsenumber{#1}%
	\pgfmathfloattomacro{\pgfmathresult}{\Fn}{\Mn}{\En}%
	\pgfmathparse{\Fn==2 ? "-" : ""}%
	\edef\Sn{\pgfmathresult}%
	\pgfmathfloatparsenumber{#2}%
	\pgfmathfloattomacro{\pgfmathresult}{\Fe}{\Me}{\Ee}%
	\pgfmathparse{int(sqrt((\Ee-\En)^2))}%
	\edef\precisionAbsEe{\pgfmathresult}%
	\pgfmathparse{int(\Ee-\En)}%
	\edef\precisionE{\pgfmathresult}%
	\pgfmathparse{\eval{\Me*10^(\precisionE)}}%
	\ifthenelse{\En=0}%
	{%
		$\Sn\pgfmathprintnumber[fixed, precision=\precisionAbsEe, zerofill]{\Mn} \pm (\pgfmathprintnumber[std, precision=0, zerofill]{#2})$%
	}%
	{%
		\ifthenelse{\En=-1}%
		{%
			\pgfmathparse{\eval{\Mn/10}}%
			\edef\Mn{\pgfmathresult}%
			\pgfmathparse{int(\precisionAbsEe+1)}%
			\edef\precisionAbsEe{\pgfmathresult}%
			$\Sn\pgfmathprintnumber[fixed, precision=\precisionAbsEe, zerofill]{\Mn} \pm (\pgfmathprintnumber[std, precision=0, zerofill]{#2})$%
		}%
		{%
			$\left(\Sn\pgfmathprintnumber[std, precision=\precisionAbsEe, zerofill]{\Mn} \pm \pgfmathprintnumber[fixed, precision=\precisionAbsEe, zerofill]{\pgfmathresult}\right)\cdot10^{\En}$%
		}%
	}%
}

\graphicspath{{figures/}}
\newacronym{PCMO}{PCMO}{praseodymium\hyp calcium\hyp manganite}
\newacronym{1D}{1D}{one\hyp dimensional}
\newacronym{2D}{2D}{two\hyp dimensional}
\newacronym{MPS}{MPS}{matrix\hyp product state}
\newacronym{MPO}{MPO}{matrix\hyp product operator}
\newacronym{SVD}{SVD}{singular\hyp value decomposition}
\newacronym{QCS}{QCS}{quantum\hyp computer simulator}
\newacronym{QC}{QC}{quantum computer}
\newacronym{FSM}{FSM}{finite\hyp state machine}
\newacronym{ACA}{ACA}{adaptive cross\hyp approximation}
\newacronym{CDW}{CDW}{charge\hyp density wave}
\newacronym{fig_CDW}{CDW}{charge\hyp density wave}
\newacronym{SDW}{SDW}{spin\hyp density wave}
\newacronym{ARPES}{ARPES}{angle-resolved photoemission spectroscopy}
\newacronym{OBC}{OBC}{open-boundary conditions}
\newacronym{PBC}{PBC}{periodic-boundary conditions}
\newacronym{TEBD}{TEBD}{time-evolution block-decimation}
\newacronym{iff}{iff}{if and only if}
\newacronym{DFT}{DFT}{density\hyp functional theory}
\newacronym{DMFT}{DMFT}{dynamical mean\hyp field theory}
\newacronym{DMRG}{DMRG}{density\hyp matrix renormalization group}
\newacronym{QMC}{QMC}{quantum Monte Carlo}
\newacronym{AIM}{AIM}{Anderson impurity model}
\newacronym{SIAM}{SIAM}{single impurity Anderson model}
\newacronym{LDA}{LDA}{local\hyp density approximation}
\newacronym{LBNL}{LBNL}{Lawrence Berkeley National Laboratory}
\newacronym{VQE}{VQE}{variational\hyp quantum eigensolver}
\newacronym{ED}{ED}{exact diagonalization}
\newacronym{QPT}{QPT}{quantum phase transition}
\newacronym{QCP}{QCP}{quantum critical point}
\newacronym{ETH}{ETH}{eigenstate thermalization hypothesis}
\newacronym{AKLT}{AKLT}{Affleck\hyp Lieb\hyp Kennedy\hyp Tasaki}
\newglossaryentry{QR}{name={QR},description={QR decomposition}}
\newacronym{TNS}{TNS}{tensor\hyp network state}
\newacronym{NN}{NN}{nearest\hyp neighbor}
\newacronym{NNN}{NNN}{next\hyp nearest\hyp neighbor}
\newacronym{TLL}{TLL}{Tomonaga-Luttinger liquid}
\newacronym{PLL}{PLL}{pair Luttinger liquid}
\newacronym{fig_PLL}{PLL}{pair Luttinger liquid}
\newacronym{FB}{FB}{flat band}
\newacronym{fig_FB}{FB}{flat band}
\newacronym{C}{C}{coexistence}
\newacronym{SU}{SU}{superfluid}
\newacronym{fig_SU}{SU}{superfluid}
\newacronym{PH}{PH}{projected Hamiltonian}
\newacronym{IP}{IP}{itinerant-paired}
\newacronym{fig_IP}{IP}{itinerant-paired}
\newacronym{QO}{QO}{quasi-order}
\newacronym{fig_QO}{QO}{quasi-order}
\newacronym{PRM}{PRM}{purely repulsive model}
\newacronym{fig_PRM}{PRM}{purely repulsive model}
\newacronym{MIM}{MIM}{mixed interaction model}
\newacronym{fig_MIM}{MIM}{mixed interaction model}
\newacronym{LL}{LL}{Luttinger liquid}
\newacronym{CFT}{CFT}{conformal field theory}

\begin{document}
\def\thetitle{Superconducting pairing from repulsive interactions of fermions in a flat-band system}
\title{\thetitle}
\author{I. Mahyaeh}
\email{iman.mahyaeh@physics.uu.se}
\affiliation{Department of Physics and Astronomy, Uppsala University, Box 516, S-751 20 Uppsala, Sweden}
\author{T. Köhler}
\affiliation{Department of Physics and Astronomy, Uppsala University, Box 516, S-751 20 Uppsala, Sweden}
\author{A. M. Black-Schaffer}
\email{annica.black-schaffer@physics.uu.se}
\thanks{equal contributor}
\affiliation{Department of Physics and Astronomy, Uppsala University, Box 516, S-751 20 Uppsala, Sweden}
\author{A. Kantian}
\email{a.kantian@hw.ac.uk}
\thanks{equal contributor}
\affiliation{SUPA, Institute of Photonics and Quantum Sciences, Heriot-Watt University, Edinburgh EH14 4AS, United Kingdom}
\affiliation{Department of Physics and Astronomy, Uppsala University, Box 516, S-751 20 Uppsala, Sweden}
\date{\today}

\begin{abstract}
Fermion systems with flat bands can boost superconductivity by enhancing the density of states at the Fermi level. We use quasiexact numerical methods to show that repulsive interactions between spinless fermions in a \gls{1D} flat-band system, the Creutz ladder, give a finite pairing energy that increases with repulsion, though charge \gls{QO} remains dominant. Adding an attractive component shifts the balance in favor of superconductivity and the interplay of two flat bands further yields a remarkable enhancement of superconductivity, well outside of known paradigms for 1D fermions.
\end{abstract}
\maketitle 

Systems with flat-band dispersions have recently attracted much theoretical and experimental attention~\cite{Bistritzer12233, wilson,TBG2, Flannigan_2020, Obs_Lieb_lat,Volovik,Volovik_2,SC_enh_01,SC_enh_02,Peotta,Annica_Thomas,TBG1,young}, because when the kinetic energy of quantum particles is near zero at some or all momenta, interactions become completely decisive for the many-body ground state. In particular, a major line of work has been devoted to the enhancement of superconductivity in electronic systems with flat bands near the Fermi level~\cite{Volovik,Volovik_2,SC_enh_01,SC_enh_02,Peotta,Annica_Thomas,TBG1,young,Mondaini18}. While significant boost to superconductivity is generally found in flat-band systems due to the diverging density of states, previous works have predominantly inserted the electron pairing, which is at the root of superconductivity, just by hand, i.e., only using an assumed effective attractive interaction between electrons. However, microscopic interactions are primarily repulsive between electrons. In non-flat-band systems, electron pairing, and subsequently superconductivity, \textit{might} emerge from repulsive interactions, but charge or magnetic orders are also closely competing, making reliable results on unconventional and high-$T_c$ superconductivity extremely challenging to obtain~\cite{LeBlanc2015}. 

How the competition between superconducting and charge or magnetic orders plays out in flat-band systems with repulsive electron interactions is an almost unexplored area, which the present work addresses. Mean-field results have shown that superconductivity is more robust than charge orders when doping away from the flat band~\cite{Annica_Thomas}, further motivating us to study the competition between different orders using the most reliable approaches possible. For this purpose, \gls{1D} systems are ideal targets, as \gls{DMRG} techniques~\cite{White92, White93, Schollwoeck05, Schollwoeck11} yield quasiexact results. It also allows us to study the effect of flat bands in \gls{1D} systems specifically, a burgeoning field in itself~\cite{Mondaini18, Katsura13, Huber13, HuberAltman10, Flannigan_2020}. In particular, flat-band systems might fall outside the widely used low-energy \gls{LL} description of 1D fermions, as the necessary linearization of the dispersion breaks down~\cite{Giamarchi}. 
 
\captionsetup[subfigure]{position=top,singlelinecheck=off,justification=raggedright} %
\begin{figure}[!t]%
	\centering%
	\tikzsetnextfilename{Creutz_ladder}%
	\begin{tikzpicture}%
		\clip (-4.,2) rectangle + (8,-2.5);
		\begin{scope}%
		[%
			node distance = 0.02,%
		]%
		\foreach \x in {-2,-1,0,+1,+2}%
		{%
			\foreach \y in {0,1}%
			{%
				\node[circle, draw,inner sep=1pt] at ($(\x*3.0,\y*1.5)$) (\x\y) {\tiny $\phantom{b_{j+2}}$};%
				\node[circle, inner sep=0] at (\x\y) (\x\y Text) %
				{%
					\tiny \ifthenelse{\y=0}%
					{%
						$b_{j\ifthenelse{\x=0}{}{\x}}$%
					}%
					{%
						$a_{j\ifthenelse{\x=0}{}{\x}}$%
					}%
				};%
			}%
		}%
		\foreach \y in {0,1}%
		{%
			\draw[very thick,dotted] (-2\y) to (-1\y);
			\draw[very thick,dotted] (+1\y) to (+2\y);
		}%
		\draw[very thick,dotted] (-20) to (-11);
		\draw[very thick,dotted] (-21) to (-10);
		\draw[very thick,dotted] (+20) to (+11);
		\draw[very thick,dotted] (+21) to (+10);
		\node at (-4,1.8) {\subfloat[\label{fig-1_creutz_ladder}]{}};
		\foreach \x [remember=\x as \lastx (initially -1)] in {-1,0,+1}%
		{%
			\ifthenelse{\x>-1}%
			{%
				\foreach \y [remember=\y as \lasty (initially 1)] in {0,1}%
				{	%
					\draw[] (\lastx\lasty) -- node[midway] (m) {} (\x\y);%
					\ifthenelse{\y=0}%
					{%
						\node[anchor=south east, inner xsep = 1.0em, inner ysep = 1.0em] at (m) {\footnotesize $-t$};%
						\node[anchor=south west, inner xsep = 1.0em, inner ysep = 1.0em] at (m) {\footnotesize $-t$};%
					}%
					{%
						\node[anchor=north east, inner xsep = 1.8em, inner ysep = 0.1em] at (m) {\footnotesize $V_2$};%
						\node[anchor=north west, inner xsep = 1.8em, inner ysep = 0.1em] at (m) {\footnotesize $V_2$};%
					}%
					\ifthenelse{\y=0}%
					{%
						\draw[]  (\lastx\y) -- %
						node[above, inner sep = 1mm] {\footnotesize $it$}%
						node[below, inner sep = 1mm] {\footnotesize $V_1$}%
						(\x\y);%
					}%
					{%
						\draw[]  (\lastx\y) -- %
						node[below, inner sep = 1mm] {\footnotesize $-it$}%
						node[above, inner sep = 1mm] {\footnotesize $V_1$}%
						(\x\y);
					}%
				}%
			}%
			{}%
		}%
		\end{scope}%
	\end{tikzpicture}%
\\[-0.75em]
	\def\pdl{3.3}
	\def\fbc{1/3}
	\def\cpll{2/5}
	
	\def\ymax{3}
	\tikzsetnextfilename{PRM_and_MIM}%
	\begin{tikzpicture}[scale=1]%
		\begin{groupplot}%
		[%
			group style = %
			{%
				group size 			=	2 by 1,%
				horizontal sep		=	1.5em,%
				vertical sep		=	0em,%
				x descriptions at	=	edge bottom,%
				y descriptions at	=	edge left,%
			},%
			width	=	0.26\textwidth,%
			height	=	0.125\textheight,%
			xmin	=	0,%
			xmax	=	1,%
			ymin	=	0,%
			ymax	=	\ymax,%
			ylabel	=	{\small $V_1/\Delta\epsilon$},%
		]%
			\nextgroupplot%
			[%
				title			=	{\acrshort{PRM}},%
				title style		=	{yshift=-0.75em},					%
				ytick			=	{0,1,2},%
				yticklabels		=	{\tiny 0, \tiny \nicefrac14, \tiny \nicefrac12},%
				xtick			=	{0,0.333,0.4,1},%
				xticklabels		=	{\tiny 0, \tiny $\frac13$, \tiny $\frac25$, \tiny 1},%
				xtick pos		=	left,%
				point meta min	=	0,%
				point meta max	=	3,%
				colormap/blackwhite,%
				axis on top,%
			]%
				\draw[] (0.333,0) to (0.333,\ymax);%
				\addplot%
				[%
					point meta	=	{y},%
					mesh,%
				] table {data/line.dat};%
				\node at (0.166,1) {\acrshort{FB}};%
				\node[text width=6em, align=center] at (0.7,1) {\acrshort{IP}\\{\tiny(charge QO)}};%
				\coordinate (n1) at (0.5,-0.5);%
				\coordinate (l1) at (-0.25,3.75);%
				\node[rotate = 90] at (0.36,1.5) {\tiny coexistence};%
			\nextgroupplot%
			[%
				title		=	{\acrshort{MIM}},%
				title style	=	{yshift=-0.75em},%
				xtick		=	{0,0.73,1},%
				xticklabels	=	{\tiny 0, \tiny $0.73$, \tiny 1},%
				xtick pos	=	left,%
				axis on top,%
			]%
				\pgfplotsset{%
					legend image code/.code={%
						\draw [#1] (0cm,-0.1cm) rectangle (0.3cm,0.1cm);%
					},%
				}%
				\node at (0.5,2.5) {phase sep};%
				\addplot %
				[%
					pattern=north west lines,%
					pattern color=colorIPphase,%
					draw=none,%
					samples=500,%
				]%
					coordinates %
					{%
						(0,0)%
						(0,2.1)%
						(0.73,2.1)%
						(0.73,2)%
						(0.6,1.0)%
						(0.5,0.5)%
						(0.5,0)%
					};%
				\addplot %
				[%
					pattern=north east lines,%
					pattern color=colorIPcharge,%
					draw=none,%
					samples=500,%
				]%
					coordinates %
					{%
						(0.5,0)%
						(0.5,0.5)%
						(0.6,1.0)%
						(0.73,2)%
						(0.73,2.1)%
						(1,2.1)%
						(1,0)%
					};%
				\addplot %
				[%
					pattern=vertical lines,%
					pattern color=gray,%
					draw=gray,%
					samples=500,%
				]%
					coordinates %
					{%
						(0,1.9)%
						(0,2.1)%
						(1,2.1)%
						(1,1.9)%
						(0,1.9)%
					};%
				\draw[] (0.73,2.0) to (0.6,1.0) to (0.5,0.5) to (0.5,0.0);%
				\addplot%
				[%
					mark	=	pentagon,%
				]%
					coordinates {(0.73,2.0) (0.6,1.0) (0.5,0.5)};%
				\coordinate (pll) at (0.5,1);%
				\coordinate (pqo)  at (0.25,0.1);%
				\coordinate (cqo)  at (0.75,0.1);%
				\coordinate (n2) at (0.5,-0.5);%
				\coordinate (l2) at (-0.05,3.75);%
				\coordinate (ip) at (1,2);
		\end{groupplot}%
		\node[fill=white,fill opacity=0.7,inner sep=0.5mm] at (pll) {\acrshort{PLL}};%
		\node[fill=white,fill opacity=0.7,inner sep=0,anchor=south] at (pqo) {\tiny (phase QO)};%
		\node[fill=white,fill opacity=0.7,inner sep=0,anchor=south] at (cqo) {\tiny (charge QO)};%
		\node at (l1) {\subfloat[\label{fig-pd_repulsive_model}]{}};%
		\node at (n1) {$n$};%
		\node at (l2) {\subfloat[\label{fig-pd_partial_attraction_model}]{}};%
		\node at (n2) {$n$};%
		\node[anchor=west,inner sep=0] at ($(ip)+(0.2,0)$) (iptext) {IP};%
		\draw[->] (iptext) to (ip);
	\end{tikzpicture}%
	\caption%
	{%
		\label{fig_1_lattice_pds}%
		\subfigref{fig-1_creutz_ladder} Creutz ladder with sites $a_j$ and $b_j$ in the $j$-th unit cell with hopping and interaction parameters given by the Hamiltonian \cref{eq-H}. %
		\subfigref{fig-pd_repulsive_model} Phase diagram of the \glsfirst{fig_PRM} with $V_1 >0$ and $V_2=0$ as a function of density $n$. %
		For $n\leq 1/3$ the physics is captured by the lower flat band  without interactions. %
		For $n > 2/5$ an \glsfirst{fig_IP} phase of fermions exists with dominant charge \glsfirst{fig_QO}. %
		In between we find a coexistence regime. %
		\subfigref{fig-pd_partial_attraction_model} Phase diagram of the \glsfirst{fig_MIM} with $V_1=-V_2>0$ with a \glsfirst{fig_PLL} for $V_1/\Delta \epsilon \lesssim 1/2 $, with phase (charge) QO for lower (higher) densities marked by three points. %
		Around $V_1/\Delta \epsilon \simeq 1/2 $ an IP with highly enhanced superconducting correlations exists for $n\leq 0.73$. Larger interactions give phase separation. %
	}%
\end{figure}
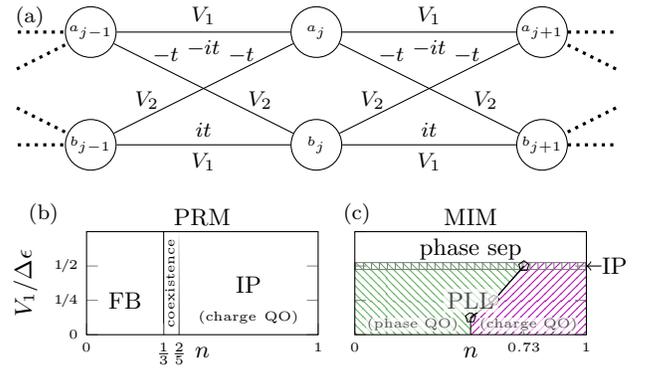

In this work, we focus on a concrete \gls{1D} flat-band fermion system, the Creutz ladder~\cite{Creutz} (depicted in ~\cref{fig_1_lattice_pds}\subref{fig-1_creutz_ladder}), as it has been realized experimentally using a parametric cavity for bosons~\cite{wilson} and an optical lattice for fermions~\cite{Kang2020}, and for which the physics of repulsively interacting bosons has recently also been investigated~\cite{Katsura13,Huber13,HuberAltman10,Flannigan_2020}. The Creutz ladder has two flat bands separated by a finite energy gap, and by employing \gls{DMRG} numerics, we obtain the many-body ground state in the presence of repulsive interactions between spinless fermions.
We find that an \gls{IP} phase of fermions exists for a wide range of densities, with a pairing energy that even increases monotonically with repulsion. At the same time, we find that pure repulsion gives a dominant charge \gls{QO}. To achieve dominant superconducting phase \gls{QO}, we tune the interactions by adding some attraction. Here, in the weak interaction regime with the physics described by a single flat band, we realize a previously proposed state of a hole-based superconductor~\cite{Maciej16}, forming a \gls{PLL}~\cite{Maciej16,Giamarchi}. Most remarkably, with increasing interaction strength, we find that the two flat bands in conjunction lead to a massive boost in superconducting correlations, far beyond known \gls{1D} paradigms.

\paragraph{Setup.---} We study interacting spinless fermions on the Creutz ladder, comprised of sites $a_j$ and $b_j$ in the $j$-th unit cell, with annihilation operators $\hat f_{a,j}$ and $\hat f_{b,j}$. The Hamiltonian we focus on is
\begin{align}%
	\label{eq-H}%
	\hat H=\sum_{j} 	& t\left(i \hat f^{\dagger}_{b,j+1} \hat f^{\nodagger}_{b,j} - i \hat f^{\dagger}_{a,j+1} \hat f^{\nodagger}_{a,j}  + \mathrm{H.c.} \right)  \nonumber \\%
-	& t\left( \hat f^{\dagger}_{b,j+1} \hat f^{\nodagger}_{a,j} + \hat f^{\dagger}_{a,j+1} \hat f^{\nodagger}_{b,j} + \mathrm{H.c.} \right)  \nonumber \\%
						+ & V_1 \left(\hat n^{\nodagger}_{b,j} \hat n^{\nodagger}_{b,j+1} + \hat n^{\nodagger}_{a,j} \hat n^{\nodagger}_{a,j+1}\right) \nonumber\\
						+ & V_2 \left( \hat n^{\nodagger}_{b,j} \hat n^{\nodagger}_{a,j+1} + \hat n^{\nodagger}_{a,j} \hat n^{\nodagger}_{b,j+1}\right)\;,%
\end{align}%
where $\hat n_{a/b,j}=\hat f^{\dagger}_{a/b,j} \hat f_{a/b,j}$, $V_1$ denotes the \gls{NN} interaction along the legs, and $V_2$ \gls{NNN} interaction along the diagonals. We set the overall energy scale by fixing $t=1$.%
The flat-band dispersion becomes explicit by rewriting \cref{eq-H} in the basis diagonalizing  the kinetic energy, using the maximally localized Wannier basis~\cite{Katsura13,Huber13,Flannigan_2020}
\begin{align}%
	\label{eq-band_operators}%
	\hat c_{\pm,j} = \frac{1}{2} \left[ \left(\hat f_{a,j} - i \hat f_{b,j} \right) \mp \left( \hat f_{a,j+1} - i \hat f_{b,j+1} \right) \right] \;.%
\end{align}%
In this basis, the kinetic energy terms give two flat bands at energies $\epsilon_\pm = \pm 2$, with a gap ${ \Delta \epsilon = 4}$. Focusing on the lower band (upper band behaves similarly) and including interactions, we arrive at the single-band projected Hamiltonian
\begin{align}%
	\label{eq-projected_H}%
	\hat H_{-}  = \sum_j 	& \epsilon_- \hat n^{\nodagger}_{-,j} + t_p \left(\hat c^{\dagger}_{-,j+1} \hat c^{\dagger}_{-,j} \hat c^{\nodagger}_{-,j} \hat c^{\nodagger}_{-,j-1} + \mathrm{H.c.} \right) \nonumber \\%
						& +  U_1 \hat n^{\nodagger}_{-,j} \hat n^{\nodagger}_{-,j+1} + U_2 \hat n^{\nodagger}_{-,j} \hat n^{\nodagger}_{-,j+2}  \;,%
\end{align}%
where $t_p=(V_1 - V_2)/8$, $U_1=2U_2=(V_1+V_2)/4$, and $\hat n^{\nodagger}_{-,j}=\hat c^{\dagger}_{-,j} \hat c^{\nodagger}_{-,j}$. %
This representation makes it explicit that in the single-band limit the physics is that of density-assisted tunneling, equivalent to tunneling of \gls{NN} pairs of fermions, with a purely interaction-set amplitude $t_p$, plus remnant short-range interactions $U_1$ and $U_2$. 

In terms of interactions, we first study a \gls{PRM}, where we set ${V_1>0,V_2=0}$, generating ${U_1,U_2>0}$ in the single-band limit. Later, we expand our focus by adding an attractive interaction $V_2<0$, resulting in $U_1=U_2=0$, which we name the \gls{MIM}. We solve the full Hamiltonian, \cref{eq-H}, numerically using two \gls{DMRG} codes~\cite{ALPS,SymMPS} to quasiexactly obtain the many-body ground states. Here, we exploit the conserved particle number, such that our results are presented as a function of density $n= N/L_x$, where $N$ is the total number of fermions and $L_x$ is the ladder's linear size, restricted to ${0 < n \leq 1}$ due to particle-hole symmetry of the Hamiltonian.
Observables, apart from gap extrapolations, are computed at ${ L_x = 100 }$ with the default bond dimension ${ \chi = 512 }$, but we ascertained convergence using up to ${ \chi = 1024 }$. 
Studying the single-band projected Hamiltonian, \cref{eq-projected_H}, alongside the quasiexact DMRG results, provides us with further insights.

To determine the characteristics of the many-body ground states, we first show that single-particle excitations are gapped and that low-lying excitations consist of fermion pairs by computing the one- and two-particle gaps ($m=1,2$) 
\begin{align}%
	\Delta_{m}(n) = E_0(n L_x+m) + E_0(n L_x - m) -2 E_0(nL_x) \;,%
\end{align}%
where $E_0(N)$ denotes the ground-state energy of $N$ fermions. These gaps are computed at different $L_x$ and then extrapolated to ${1 / L_x \rightarrow 0}$. %
We also obtain important information by focusing on the lower band. Here, we study both the phase (ph) and charge-density (ch) correlation functions
\begin{align}%
	\label{eq_g_m_2} G_{-,\mathrm{ph}}(r) &= |\braket{\hat c^{\dagger}_{-,j} \hat c^{\dagger}_{-,j+1} \hat c^{\nodagger}_{-,j+1+r} \hat c^{\nodagger}_{-,j+r}}| \;, \\%
 \label{eq_g_m_nn} G_{-,\mathrm{ch}}(r)&= |\braket{\hat n^{\nodagger}_{-,j} \hat n^{\nodagger}_{-,j+r}}- \braket{\hat n^{\nodagger}_{-,j}}\braket{\hat n^{\nodagger}_{-,j+r}}|\; ,%
\end{align} %
particularly extracting their power-law decay exponents $\alpha_{-,\mathrm{ph/ch}}$ to determine if phase or charge \gls{QO} is slowest decaying and thus dominant. We also compute the single-particle density matrix of the lower band
\begin{align}%
	\label{eq_g_m_1} G_{-}(r) &= |\braket{\hat c^{\dagger}_{-,j} \hat c^{\nodagger}_{-,j+r}}|,%
\end{align}%
to verify that its single-particle excitations are gapped. As a further marker of fermion pairing~\cite{RA, Gotta_Mazza_PRL, Gotta_Mazza_PRR, Gotta_Mazza_arXiv}, we calculate the Fourier transform of the fluctuations of the local density, ${ \mathcal{FT}[\delta \langle \hat n_{-,j}\rangle ] }$. 
In the case of strong interactions, we additionally find it useful to analyze the density-density correlation function across the diagonal links of the ladder in the original two-band basis
\begin{align}%
	\label{eq_g_diag} G_{\mathrm{diag}}(r)  &= \braket{\hat n^{\nodagger}_{b,j} \hat n^{\nodagger}_{a,j+1} \hat n^{\nodagger}_{b,j+r} \hat n^{\nodagger}_{a,j+1+r}} \nonumber \\
& - \braket{\hat n^{\nodagger}_{b,j} \hat n^{\nodagger}_{a,j+1}} \braket{\hat n^{\nodagger}_{b,j+r} \hat n^{\nodagger}_{a,j+1+r}}\; .%
\end{align}%
Finally, for complementary information, we obtain the bipartite entanglement entropy $S(l) = -\mathrm{Tr} \left[ \rho(l) \ln \rho(l)   \right]$ for $ 1 \leq l < 2 L_x$, where $\rho(l)$ is the reduced density matrix of the first $l$ sites~\cite{reduced_rho}. %

\captionsetup[subfigure]{position=top,singlelinecheck=off,justification=raggedright} %
\begin{figure}[!t]%
	\centering%
	\tikzsetnextfilename{single_particle_ns}%
	\begin{tikzpicture}%
		\def\nfacs{{10,12,14,16}}%
		\begin{groupplot}%
			[%
			group style = %
			{%
				group size 			=	1 by 3,%
			},%
			height			= 	0.175\textheight,%
			width 			= 	0.475\textwidth,%
			]%
			\nextgroupplot%
			[%
			xlabel			=	{$V_1/\Delta\epsilon$},%
			xlabel style	=	{yshift = 0.7em},
			ylabel			=	{$\Delta_1$},%
			ymin			=	0,%
			ymax			=	0.16,%
			ytick			=	{0, 0.1},%
			yticklabels		=	{0, 0.1},%
			xtick			=	{0, 0.0625, 0.125, 0.25, 0.5},%
			xticklabels		=	{0, $\nicefrac{1}{16}$, $\nicefrac18$, $\nicefrac14$, $\nicefrac12$},%
			legend columns	=	2,%
			legend transposed,%
			legend style	=%
			{%
				font=\scriptsize,%
				draw=none,%
				fill opacity=0.8,%
				at={(0.975,0.975)},%
				anchor=north east%
			},%
			]%
			\pgfmathtruncatemacro{\dimension}{dim(\nfacs)-1}%
			\expandafter\pgfplotsinvokeforeach\expandafter{0,...,\dimension}%
			{%
					\pgfmathparse{\nfacs[#1]/20}\pgfmathprintnumberto{\pgfmathresult}{\n}%
					\pgfmathtruncatemacro{\nfac}{\nfacs[#1]}%
					\pgfmathsetmacro{\marker}{\gpmarkers[mod(#1,dim(\gpmarkers))]}%
					\pgfmathsetmacro{\cl}{\gpcolors[mod(#1,dim(\gpcolors))]}%
					\edef\cmd%
					{%
						\noexpand\addplot%
						[%
							color	= \cl,%
							mark 	= \marker,%
						]%
							table%
							[%
								x expr = \noexpand\thisrowno{1}/4.0,%
								y expr = \noexpand\thisrowno{2},%
							]%
							{data/SymMPS/spinlessFermions/single_particle_gap_summary_Nfac_\nfac.dat};%
						\noexpand\addlegendentry{$n=\n$};%
					}%
					\cmd%
				}%
			\coordinate (insetPosition) at (rel axis cs:0.095,0.53);%
			\coordinate (l1) at (rel axis cs:-0.15,1);%
			\nextgroupplot%
			[%
			xlabel	=	{$n$},%
			ylabel			=	{Decay exponent},%
			ylabel style	=	{yshift=-0.5em,},%
			legend style = %
			{%
				font	=	\scriptsize,%
				draw	=	none,%
				fill	=	none,%
				at		=	{(rel axis cs:0.05,0.95)},%
				anchor	=	north west,%
			},%
			]%
				\addplot%
				[%
					color	=	colorIPphase,%
					thick,%
					mark	=	triangle*,%
				]%
					table {data/MAQUIS/m1_alpha_n.dat};%
				\addlegendentry{$\alpha_{-,\mathrm{ph}}$}%
				\addplot%
				[%
					color	=	colorIPcharge,%
					thick,%
					mark	=	*%
				]%
					table {data/MAQUIS/m1_nn_exp_n.dat};%
				\addlegendentry{$\alpha_{-,\mathrm{ch}}$}%
				\coordinate (l2) at (rel axis cs:-0.15,1);%
		\end{groupplot}%
		\node at (l1) {\subfloat[\label{fig-m1_single_gap}]{}};%
		\node at (l2) {\subfloat[\label{fig-m1_c}]{}};%
		%
		\def\Vzeros{0p25,0p5,1,2}%
		\begin{axis}%
			[%
			at				=	{(insetPosition)},%
			width			=	0.225\textwidth,%
			height			=	0.1125\textheight,%
			xmin			=	0.4,%
			xmax			=	0.9,%
			ymin			=	0,%
			ymax			=	0.16,%
			xlabel			=	{$n$},%
			ticklabel style	=	{font=\tiny},%
			x label style	=%
			{%
				font	=	{\footnotesize},%
				at		=	{(axis description cs:0.6,-0.05)},%
				anchor	=	north,%
			},%
			ylabel			=	{$\Delta_1$},%
			y label style	=%
			{%
				font	=	{\footnotesize},%
				at		=	{(axis description cs:-0.025,0.35)},%
				anchor	=	south,%
			},%
			ytick			=	{0, 0.1},%
			]%
			\expandafter\pgfplotsinvokeforeach\expandafter{\Vzeros}%
			{%
				\StrSubstitute{#1}{p}{.}[\Vzero]%
				\pgfmathtruncatemacro{\cl}{int(100/3*ln(\Vzero*4)/ln(2))}%
				\edef\cmd%
				{%
					\noexpand\addplot%
					[
					color	= 	red!\cl!blue,%
					mark 	=	x,%
					error bars/.cd,%
					y dir	=	both,%
					y explicit,%
					]%
					table%
					[%
					x expr			=	\noexpand\thisrowno{0}/20,%
					y expr			=	\noexpand\thisrowno{2},%
					y error expr	=	\noexpand\thisrowno{3},%
					]%
						{data/SymMPS/spinlessFermions/single_particle_gap_summary_Vzero_#1.dat};%
				}%
				\cmd%
			}%
			\draw [->] (axis cs:0.75,0.005)--(axis cs:0.8,0.135) node[anchor=west,inner sep=0] {\tiny $V_1$};%
		\end{axis}%
	\end{tikzpicture}%
	\caption%
	{PRM results. %
		\label{fig-on_model_1}%
		\subfigref{fig-m1_single_gap} Single-particle gap $\Delta_1$ as a function of interaction $V_1$ for various densities $n$ in the \gls{fig_IP} phase. %
		Inset shows same data with the role of $n$ and $V_1$ interchanged, with $V_1$ increasing in direction of the arrow. %
		\subfigref{fig-m1_c} Power-law decay exponents $\alpha_{-,\mathrm{ph(ch)}}$ for phase (charge) correlations as a function of density $n$ in the \gls{fig_IP} phase at $V_1= \Delta \epsilon/2$. 
	}%
\end{figure}
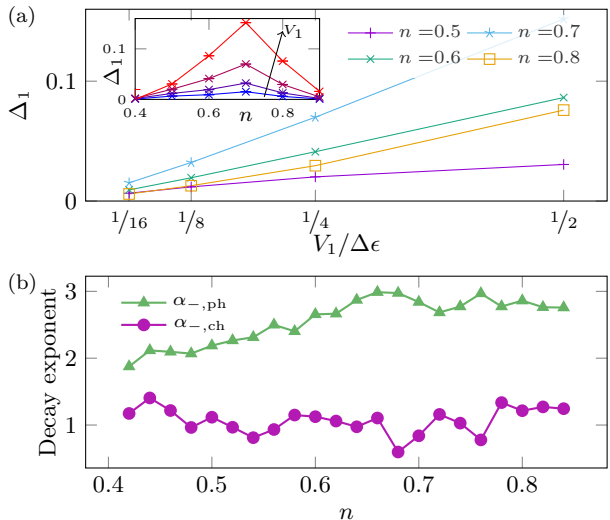%

\paragraph{PRM results.---}%
The complete many-body phase diagram for the purely repulsive model (PRM) with ${ V_1 > 0}$ and ${ V_2 = 0 }$ is summarized in~\cref{fig_1_lattice_pds}\subref{fig-pd_repulsive_model}, with a large part of the phase diagram covered by an \gls{IP} phase. This is consistent with the single-band Hamiltonian, \cref{eq-projected_H}, displaying explicit pairing physics through the non-zero $t_p$ term.

Specifically, we find that by obtaining the ground state of the full Hamiltonian~\cref{eq-H}, densities ${ n \geq 2/5}$ give ${ \Delta_1 > 0 }$ and $G_-(r)$ decaying exponentially with distance $r$. Thus, single-particle excitations play no role, in strong contrast to repulsively interacting spinless fermions on ladders with non-flat bands~\cite{Giamarchi}. %
Even more remarkably, the pairing energy $\Delta_1$ increases approximately linearly with the repulsion $V_1$ across the entire range ${\Delta \epsilon / 16 \leq V_1 \leq \Delta \epsilon / 2}$, as summarized in~\cref{fig-on_model_1}\subref{fig-m1_single_gap}. This is also in contrast to the behavior of other 1D spinless fermion systems, such as the two-leg Hubbard ladder~\cite{Giamarchi}. 
We further find ${ \Delta_2 = 0 }$ in the thermodynamic limit and that the peak of ${ \mathcal{FT}[\delta \langle \hat n_{-,j}\rangle ] }$ occurs at momentum ${ k= 2 \pi (n/2) }$, indicating that the low-energy degrees of freedom are effectively halved and thus those of fermion pairs. %
These results clearly indicate the existence of an \gls{IP} phase with fermion pairs, and not any other higher-order bound states, at densities ${ n \geq 2/5}$. %

Within the \gls{IP} phase, we find that the strong repulsion results in a dominant charge \gls{QO} in the lower band for all ${n \geq 2/5}$ . This is quantified in~\cref{fig-on_model_1}\subref{fig-m1_c}, where we plot the power-law decay exponents $\alpha_{-,\mathrm{ph(ch)}}$ for phase (charge) correlations and find the charge-density correlations decaying slowest. %
These results are obtained at ${ V_1 = \Delta \epsilon / 2 }$, but are virtually identical at lower $V_1$ values. As a consequence, for interactions up to ${ V_1 = \Delta \epsilon / 2 }$, the only energy scale in the system is $V_1$, as also evident in the projected Hamiltonian~\cref{eq-projected_H} and in the linear scaling of $\Delta_1$ in~\cref{fig-on_model_1}\subref{fig-m1_single_gap}. Spot checks at ${ V_1 =\Delta \epsilon }$ show strong deviations from these lower $V_1$ results (not included in \cref{fig_1_lattice_pds}\subref{fig-pd_repulsive_model}). %

To check whether the \gls{IP} phase is a \gls{LL} described by a \gls{CFT} with central charge $c=1$~\cite{CFT_book,Giamarchi}, we study the entanglement entropy. %
We find significant fluctuations, which cannot be captured by the Calabrese-Cardy formula~\cite{CC_2004,CC_2009,EE_osc}. %
Still, performing a naive fit yields a central charge either with a large uncertainty or clearly different from $c=1$. Thus, we cannot identify the IP phase as a PLL (see the Supplemental Material (SM)~\cite{supp_mat} for details).%

Complementing the \gls{IP} phase, we find for lower densities a fully or partially localized flat-band behavior that is not sensitive to the repulsive interactions. At ${ n \leq 1/3 }$, we find that the many-body ground state is completely localized. We ascertain this by running ground-state searches with multiple different initial conditions, falling into three categories: (1) initially localized particles, (2) random initial matrix-product states, and (3) adiabatic transition of Hamiltonian parameters from a closed ring to the Creutz ladder. All three approaches result in different, but completely localized states with the \textit{same} Wannier orbital energy $E_0= n L_x \epsilon_-$ for ${ n \leq 1/3 }$, while all three yield the same state for ${ n > 2/5 }$~\cite{convergence_sw}. 
This numerical finding is in agreement with the analytical observation of the projected Hamiltonian~\cref{eq-projected_H} that a pair of adjacent fermions with center-of-mass momentum $k$ has an effective dispersion ${ U_1 - 2 t_p \cos k }$. Thus, with ${ U_1 = 2 t_p }$ for the \gls{PRM}, a pair always adds a positive amount of energy to the system. For densities ${ n < 1/3 }$, it is possible to keep this kinetic energy cost to zero by not creating any fermion pairs, while simultaneously making the remnant interaction $U_2$ zero by putting at most one fermion in every three unit cells. We therefore designate this phase as flat band in  \cref{fig_1_lattice_pds}\subref{fig-pd_repulsive_model}. 

Finally, at $n=1/3$, our numerics show a first-order transition into a coexistence regime between the flat band and the \gls{IP} phases for {$1/3 < n < 2/5$}. At these densities, placing two fermions adjacently cannot be fully avoided, and, as a consequence, it is favorable to create some itinerant pairs. At the same time, low-density patches can still exist intermittently, where both repulsive interactions and pair formation are avoided, resulting in fully localized fermions, overall yielding a mixture of IP and FB phases.

\captionsetup[subfigure]{position=top,singlelinecheck=off,justification=raggedright} %
\begin{figure}[!t]%
	\centering%
	\tikzsetnextfilename{FullMIMcompare}%
	\begin{tikzpicture}%
		\begin{groupplot}%
		[%
			group style = %
			{%
				group size 			=	1 by 3,%
			},%
			height			= 	0.175\textheight,%
			width 			= 	0.475\textwidth,%
		]%
			\nextgroupplot%
			[%
				xlabel			=	{$n$},%
				ylabel			=	{Decay exponent},%
				ylabel style	=	{yshift=-0.5em,},%
				legend columns	=	2,%
				legend style	=	%
			{%
				at		=	{(0.5,0.75)},%
				anchor	=	south,%
				draw=none,%
			},%
			title style		=%
			{%
				at		=	{(-0.1,0.9)},%
				anchor	=	north,%
				draw,%
			},%
			]%
			\addplot%
			[%
			color	=	colorIPphase,%
			mark	=	triangle*,%
			thick,%
			]%
			table {data/MAQUIS/m2_alpha_n.dat};%
			\addlegendentry{$\alpha_{-,\mathrm{ph}}$};%
			\addplot%
				[%
					color	=	colorIPcharge,%
			mark	=	*,%
			thick,%
				]%
			table {data/MAQUIS/m2_nn_exp_n.dat};%
			\addlegendentry{$\alpha_{-,\mathrm{ch}}$};%
			\node at (rel axis cs:-0.15,1) (l1) {};%
			\nextgroupplot%
			[%
				xlabel			=	{$l$},%
				ylabel			=	{$S(l)$},%
				legend style	=	%
				{%
					at		=	{(0.5,0.05)},%
					anchor	=	south,%
					draw=none,%
				},%
			]%
				\addplot%
				[%
					color		=	black,%
					mark		=	o,%
					mark size	=	1pt,%
				]%
					table {data/MAQUIS/vN_entropy_100_2_512_36.txt};%
				\addlegendentry{\tiny $V_1 =\Delta \epsilon/8$}%
				\addplot%
				[%
					color		=	magenta,%
					mark		=	x,%
					mark size	=	1pt,%
				]%
					table%
					[%
						x expr = \coordindex+1,%
						y expr = \thisrowno{0}*ln(2),%
					]%
						{data/SymMPS/spinlessFermions/Lx_100/N_36/V0_1/V1_m1/V2_1/V3_0/chimax_512/EE.symmps_tikz}; 
				\addlegendentry{\tiny $V_1 =\Delta \epsilon/4$}%
				\addplot%
				[%
					color		=	cyan,%
					mark		=	square,%
					mark size	=	1pt,%
				]%
					table {data/MAQUIS/vN_entropy_100_8_512_36.txt};%
				\addlegendentry{\tiny $V_1=\Delta \epsilon/2$}%
				\node at (rel axis cs:-0.15,1) (l2) {};%

			\nextgroupplot%
			[%
				xlabel			=	{$r$},%
				ylabel			=	{$G_\mathrm{diag}(r)$},%
				legend style	=	%
				{%
					at		=	{(0.5,0.025)},%
					anchor	=	south east,%
					draw=none,%
					fill=none,
				},%
				ymin			=	-3e-3,
				ymax			=	1e-3,
			]%
				\addplot%
				[%
					color		=	black,%
					mark		=	o,%
					mark size	=	1pt,%
				]%
					table
					[
						restrict x to domain	=	1:21,%
						x expr					=	\thisrowno{0}/2.0,%
						y expr					=	\thisrowno{1},%
					] 	
						{data/SymMPS/spinlessFermions/Lx_100/N_36/V0_0p5/V1_m0p5/V2_0p5/V3_0/chimax_512/i_96/j_99/connected_correlator.dat};%
				\addlegendentry{\tiny $V_1 =\Delta \epsilon/8$};%
				\addplot%
				[%
					color		=	magenta,%
					mark		=	x,%
					mark size	=	1pt,%
				]%
					table
					[
						restrict x to domain	=	1:21,%
						x expr					=	\thisrowno{0}/2.0,%
						y expr					=	\thisrowno{1},%
					] 	
					{data/SymMPS/spinlessFermions/Lx_100/N_36/V0_1/V1_m1/V2_1/V3_0/chimax_512/i_96/j_99/connected_correlator.dat};%
				\addlegendentry{\tiny $V_1 =\Delta \epsilon/4$};%
				\addplot%
				[%
					color		=	cyan,%
					mark		=	square,%
					mark size	=	1pt,%
				]%
 					table
 					[
 						restrict x to domain	=	1:21,%
 						x expr					=	\thisrowno{0}/2.0,%
 						y expr					=	\thisrowno{1},%
 					] 	
 						{data/SymMPS/spinlessFermions/Lx_100/N_36/V0_2/V1_m2/V2_2/i_96/j_99/connected_correlator.dat};%
 				\addlegendentry{\tiny $V_1=\Delta \epsilon/2$};%
 				\coordinate (inset) at (axis description cs:0.95,0.15);%
				\coordinate (l3) at (rel axis cs:-0.15,1);%
		\end{groupplot}%
		\begin{axis}
		[
			at		=	(inset),
			anchor	=	south east,
			width	=	0.21\textwidth,
			height	=	0.11\textheight,
			xticklabel style	=	{font=\tiny},
			yticklabel style	=	{font=\tiny},
			xlabel style	=	{font=\tiny, yshift=0.6em,},
			ylabel style	=	{font=\tiny, yshift=-0.5em,},
			xlabel			=	{$r$},%
			ylabel			=	{$G_\mathrm{diag.}(r)$},%
		]
		\addplot[densely dotted, domain = 0:7] {0};
			\addplot%
			[%
				color		=	black,%
				mark		=	o,%
				mark size	=	1pt,%
			]%
				table
				[
					restrict x to domain	=	1:6,%
					x expr					=	\thisrowno{0}/2.0,%
					y expr					=	\thisrowno{1},%
				] 	
					{data/SymMPS/spinlessFermions/Lx_100/N_36/V0_0p5/V1_m0p5/V2_0p5/V3_0/chimax_512/i_96/j_99/connected_correlator.dat};%
			\addplot%
			[%
				color		=	magenta,%
				mark		=	x,%
				mark size	=	1pt,%
			]%
				table
				[
					restrict x to domain	=	1:6,%
					x expr					=	\thisrowno{0}/2.0,%
					y expr					=	\thisrowno{1},%
				] 	
					{data/SymMPS/spinlessFermions/Lx_100/N_36/V0_1/V1_m1/V2_1/V3_0/chimax_512/i_96/j_99/connected_correlator.dat};%
			\addplot%
			[%
				color		=	cyan,%
				mark		=	square,%
				mark size	=	1pt,%
			]%
				table
				[
					restrict x to domain	=	1:6,%
					x expr					=	\thisrowno{0}/2.0,%
					y expr					=	\thisrowno{1},%
				] 	
				{data/SymMPS/spinlessFermions/Lx_100/N_36/V0_2/V1_m2/V2_2/i_96/j_99/connected_correlator.dat};%
		\end{axis}
		\node at (l1) {\subfloat[\label{fig-MIM_exponents}]{}};%
		\node at (l2) {\subfloat[\label{fig-MIM_EE}]{}};%
		\node at (l3) {\subfloat[\label{fig-MIM_G_diag}]{}};%
	
	\end{tikzpicture}
	\caption%
	{MIM results. %
		\label{fig-on_model_2}%
	\subfigref{fig-MIM_exponents}Power-law decay exponents $\alpha_{-,\mathrm{ph(ch)}}$ for phase (charge) correlations as a function of density $n$ at $V_1=\Delta \epsilon/2$. %
		\subfigref{fig-MIM_EE} Entanglement entropy $S(l)$ as a function of subsystem size $l$ at density $n=0.36$ for different $V_1$. 
		\subfigref{fig-MIM_G_diag} Diagonal correlation function $G_{\mathrm{diag}}(r)$ as a function of distance $r$ at density $n=0.36$ for different $V_1$. %
		Inset shows a zoom-in at small $r$. 
	}%
\end{figure}
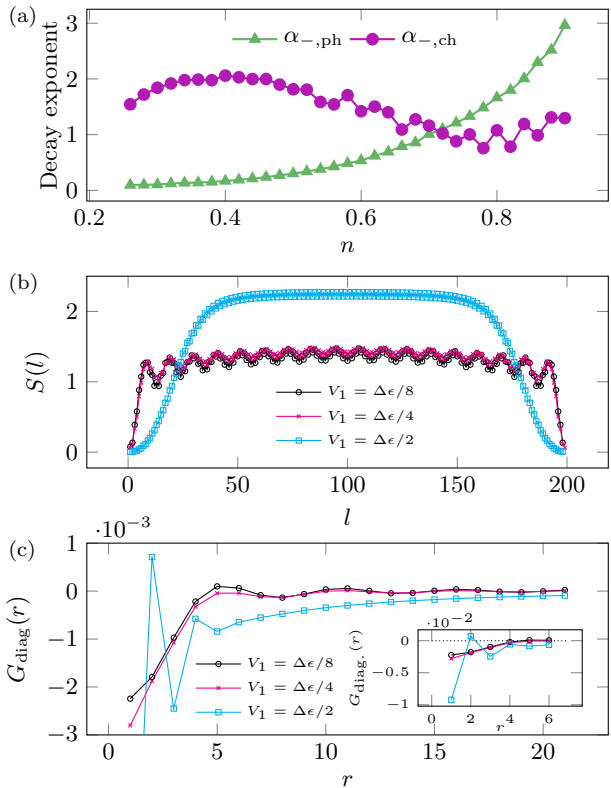%

\paragraph{MIM results.---}%
Since we find an \gls{IP} phase with dominant charge \gls{QO} within the \gls{PRM}, we next seek to enhance the superconducting correlations by reducing the remnant single-band interactions $U_{1,2}$. We do this by additionally switching on an attractive interaction $V_2$ across the diagonals of the Creutz ladder. In particular we investigate the case where $U_{1,2}$ are tuned to zero by setting ${V_2=-V_1}$, which we designate as the mixed interaction model (MIM). The complete many-body phase diagram is summarized in \cref{fig_1_lattice_pds}\subref{fig-pd_partial_attraction_model}, capturing the behavior for both $V_{1}$  below as well as when comparable with the gap $\Delta \epsilon$.

Starting with small interactions, we note that for ${ V_2 = -V_1}$, only the pair-hopping survives in the single-band projected Hamiltonian~\cref{eq-projected_H}. We additionally note that by performing a particle-hole transformation, ${\hat{c}_{-,j} \rightarrow \hat{c}^\dagger_{-,j}}$,~\cref{eq-projected_H} reduces to a model recently solved exactly by mapping to a system of hard-core bosons representing two-fermion composites~\cite{Maciej16}. 
For ${ V_1 = \Delta \epsilon / 8, \Delta \epsilon / 4}$, we find that the ground-state energies for the full system~(\cref{eq-H}) numerically agree very well with the energies of this analytical solution at all densities ${0 < n \leq 1}$ (see SM~\cite{supp_mat}). 
Furthermore, based on the exact solution~\cite{Maciej16} and its \gls{LL} description~\cite{Giamarchi}, ${G_{-,\mathrm{ph}}(r)}$ is expected to scale as $\sim r^{-1/(2K)}$. Here $K$ is the Luttinger parameter, equal to $1$ in the limit of $n \rightarrow 0$ and approaching $1/4$ as $n \rightarrow 1$, with $K \geq 1/2$ giving superconducting phase \gls{QO} \cite{Giamarchi}. Our numerics matches this behavior very closely, finding dominant superconducting \gls{QO} for all densities where ${ K \geq 1/2 }$, as marked in \cref{fig_1_lattice_pds}\subref{fig-pd_partial_attraction_model}. %
In addition, we find a finite value for $\Delta_1$, an exponential decay of $G_{-}(r)$, and a peak of ${ \mathcal{FT}[\delta \langle \hat n_{-,j}\rangle ] }$ at momentum $k= 2 \pi (n/2)$, all further confirming that the \gls{MIM} realizes a \gls{PLL} in the small interaction, single-band limit.

The behavior changes drastically in a regime around ${ V_1 \simeq \Delta \epsilon / 2 }$ where we start seeing the upper flat band affecting the physics. Our DMRG numerics find the occupation of the upper band, previously reaching at most 1.3\% and typically much less, now boosted to 3.5-8.5\%. Most interestingly, we find for ${ n \leq 0.73 }$ both a strong enhancement of the superconducting correlations and the effective physics being very different from the \gls{PLL} found at lower $V_{1}$. For example, while still decaying largely with a power law, $G_{-,\mathrm{ph}}(r)$ now shows an extremely slow decay, especially at smaller densities as illustrated in~\cref{fig-on_model_2}\subref{fig-MIM_exponents} (see SM for $V_1$ dependence ~\cite{supp_mat}). This decay is much slower than typically found for interacting fermions in 1D~\cite{Giamarchi}. We also find the entanglement entropy to be very different compared to smaller $V_{1}$, as shown in \cref{fig-on_model_2}\subref{fig-MIM_EE}, saturating to a constant value inside the bulk. Moreover, for ${n \leq 0.47 }$ we find no peak at finite momentum in ${ \mathcal{FT}[\delta \langle \hat n_{-,j}\rangle ] }$ due to complete vanishing of real-space density fluctuations. A peak at $k= 2 \pi (n/2) $, indicating pairing, only appears suddenly  for ${ n > 0.47 }$. 
At the same time, we find ${ \Delta_1 > 0 }$ as expected, and ${ \Delta_2 = 0 }$, showing that the attractive component of the interaction does not give rise to multi-particle bound states beyond pairs.

To better understand the ground state in the  regime around ${V_1\simeq \Delta \epsilon / 2 }$, we analyze the correlation function $G_{\mathrm{diag}}(r)$, as illustrated in~\cref{fig-on_model_2}\subref{fig-MIM_G_diag}. While we find the expected short-range correlation hole for composite hard-core bosons for ${ V_1 = \Delta \epsilon / 8,\Delta \epsilon / 4 }$, the behavior at ${ V_1 = \Delta \epsilon / 2 }$ is very different, with a much deeper and wider correlation hole, explaining why we see no higher-order bound states beyond pairs. There is now also a positive correlation for finding pairs at distance $r=2$ which can explain the major boost in superconducting correlations. We interpret these results as the upper flat band providing a strong additional short-range attraction for pairs in the lower flat band through virtual processes, rendering a substantial boost to superconductivity, well beyond the known PLL regime for pairing in 1D fermion systems. 

While superconducting correlations are boosted for ${ V_1\simeq \Delta \epsilon / 2 }$ in a broad density range, we eventually find charge \gls{QO} dominating for ${ n > 0.73 }$. 
In this high-density regime the entanglement entropy approaches the Calabrese-Cardy pattern, although never quite obtaining it, pointing to significant remnants of non-\gls{LL} physics.
Finally, for strong interactions ${ V_1 > \Delta \epsilon / 2 }$, we find phase separation where also higher-order bound states are formed.

\paragraph{Discussion.---} Using quasiexact many-body methods, we show how flat bands give rise to fermion pairing from purely repulsive interactions, with pairing strength even increasing with repulsion. Although we find charge-density \gls{QO} ultimately dominating for pure repulsion, it would be pre-mature to conclude that 2D or 3D arrays formed from Creutz ladders  will necessarily have a charged-ordered ground state, as weak inter-ladder coupling could help stabilize the superconducting phase. Our work also highlights the great potential of multiple flat bands to strongly stabilize superconductivity, establishing a regime beyond known paradigms for pairing in 1D systems. As Creutz ladders have been realized with neutral ultracold fermionic atoms via optical potentials~\cite{Kang2020}, our results might be tested experimentally, e.g., using dipolar ultracold gases in the near future.

\paragraph{Acknowledgments--}
A.B.-S. and A.K. contributed equally to this work: A.B.-S. provided the initial ideas, supervision, and the major share of resources, A.K. delineated the final project, provided supervision, and the minor share of resources. We would like to thank Lorenzo Gotta, Leonardo Mazza, and Paola Ruggiero for helpful discussions. %
I.M. and A.B.-S. acknowledge financial support from the Knut and Alice Wallenberg Foundation through the Wallenberg Academy Fellows program and the Swedish Research Council (Vetenskapsr\aa det Grant No.~2018-03488). %
This work has received funding through an ERC Starting Grant from the European Union's Horizon 2020 research and innovation programme under Grant agreement No.~758935. The computations were enabled by resources provided by the Swedish National Infrastructure for Computing (SNIC) at KTH Parallelldatorcentrum (PDC) and Uppsala Multidisciplinary Center for Advanced Computational Science (UPPMAX) partially funded by the Swedish
Research Council through Grant No.~2018-05973.

%

\bibliographystyle{unsrturl}
\bibliography{refs}%

\begin{thebibliography}{10}
\bibitem{reduced_rho}
If the density matrix $\rho$ describes a system with two subsystems $A$ and
  $B$, the reduced density matrix of the subsystem $A$, $\rho_A$, is calculated
  by tracing over all the degrees of freedom of the subsystem $B$, namely
  $\rho_A = \mathrm{Tr_B} \rho$. This is easily done in DMRG since all the
  Schmidt decomposition eigenvalues are known for all links.
\end{thebibliography}

\clearpage
\appendix%
\onecolumngrid%
\begin{center}%
	\textbf{\large Supplemental Materials: \thetitle}%
\end{center}%
\setcounter{equation}{0}%
\setcounter{figure}{0}%
\setcounter{table}{0}%
\setcounter{page}{1}%
\makeatletter%
\renewcommand{\theequation}{S\arabic{equation}}%
\renewcommand{\thefigure}{S\arabic{figure}}%
\renewcommand{\bibnumfmt}[1]{[S#1]}%
\onecolumngrid
\glsresetall%
\section{PRM: Entanglement entropy and central charge}%
As stated in the main text, the entanglement entropy pattern of the \gls{PRM} in the \gls{IP} phase is quite different from what is obtained from the Calabrese-Cardy formula with a central charge $c=1$. %
We use the Calabrese-Cardy formula \cite{CC_2004,CC_2009}
\begin{equation}
\label{CCansatz}
S(l) = \frac{c}{6} \ln \left( \sin \left( \frac{\pi l}{L} \right)  \right) + S_0 \ ,
\end{equation} 
to extract the central charge $c$ from our data. Here, $l$ is the length of the subsystem, $L=2L_x$ is the size of the system, and $S_0$ is a model-dependent constant. For the purpose of fitting, we also sometimes use an offset $l_0$, such that $l \in \left[ 1+l_0, Lx \right]$.

\begin{figure}[!h]%
	\centering%
	\tikzsetnextfilename{EntanglementEntropies}
	\begin{tikzpicture}%
		\begin{groupplot}%
		[%
			group style = %
			{%
				group size 			=	2 by 2,%
				horizontal sep		=	1.5em,%
				vertical sep		=	1.2em,%
				horizontal sep		=	1.2em,%
				x descriptions at	=	edge bottom,%
				y descriptions at	=	edge left,%
			},%
			height			= 	0.2\textheight,%
			width 			= 	0.5\textwidth,%
			xlabel			=	{$l$},%
			ylabel			=	{$S(l)$},%
			ymin			=	0,%
			ymax			=	1.6,%
			xmin			=	1,%
			xmax			=	199,%
		]%
			\def\ns{50,60,70,80}%
			\expandafter\pgfplotsinvokeforeach\expandafter{\ns}%
			{%
				\pgfmathtruncatemacro{\N}{int(#1)}%
				\pgfmathparse{\N/100}%
				\pgfmathprintnumberto[precision=2]{\pgfmathresult}{\n}%
				\ifthenelse{\boolean{useShellEscape}}
				{
					\immediate\write18{tail -n1 data/SymMPS/spinlessFermions/Lx_100/N_\N/V0_2/V1_0/V2_2/EE.symmps | tr ' ' '\string\n' | head -n-1 | awk '{print NR, $1*log(2)}' > data/SymMPS/spinlessFermions/Lx_100/N_\N/V0_2/V1_0/V2_2/EE.symmps.tikzready}
				}{}%
				\edef\cmd%
				{%
					\noexpand\nextgroupplot%
						\noexpand\addplot%
						[%
							mark=o,%
							mark size=1pt%
						]%
							table%
							[%
								x expr	=	\noexpand\thisrowno{0},%
								y expr	=	\noexpand\thisrowno{1},%
							]%
							{data/SymMPS/spinlessFermions/Lx_100/N_\N/V0_2/V1_0/V2_2/EE.symmps.tikzready};%
						\noexpand\node[anchor=south west] at (axis cs:5,0) {$n=$ $\n$};%
						\noexpand\ifthenelse{\noexpand\boolean{useShellEscape}}%
						{
							\noexpand\addplot%
							[]%
								gnuplot%
								[%
								 	raw gnuplot,%
								]%
								{%
									set fit quiet;%
									set fit errorvariables;%
									c=0.1;%
									S=0.1;%
									L=200;%
									f(x)=c/6.0*log(L*sin((pi*x)/L))+S;%
									fit [*:*]f(x) "<tail -n1 data/SymMPS/spinlessFermions/Lx_100/N_\N/V0_2/V1_0/V2_2/EE.symmps | tr ' ' '\string\n'" u ($0+1):($1*log(2)) via c,S;%
									set print "data/SymMPS/spinlessFermions/Lx_100/N_\N/V0_2/V1_0/V2_2/EE.fit";%
									print c,c_err,S,S_err;%
								};%
						}{}%
						\noexpand\pgfplotstableread[header=false]{data/SymMPS/spinlessFermions/Lx_100/N_\N/V0_2/V1_0/V2_2/EE.fit}{\noexpand\EEfit}%
						\noexpand\pgfplotstablegetelem{0}{0}\noexpand\of\noexpand\EEfit%
						\noexpand\pgfmathsetmacro{\noexpand\EEfitC}{\noexpand\pgfplotsretval}%
						\noexpand\pgfplotstablegetelem{0}{1}\noexpand\of\noexpand\EEfit%
						\noexpand\pgfmathsetmacro{\noexpand\EEfitCerr}{\noexpand\pgfplotsretval}%
						\noexpand\pgfplotstablegetelem{0}{2}\noexpand\of\noexpand\EEfit%
						\noexpand\pgfmathsetmacro{\noexpand\EEfitS}{\noexpand\pgfplotsretval}%
						\noexpand\addplot%
						[%
							smooth,%
							blue,%
							domain = 0.5:199.5,%
							samples=400,
						]%
							{ccFct(\noexpand\EEfitC, 200.0, \noexpand\EEfitS)};%
						\noexpand\node[anchor=south] at (axis cs:100,0) {$c=$\noexpand\printpgfnumberwitherror{\noexpand\EEfitC}{\noexpand\EEfitCerr}};%
				}\cmd%
			}%
		\end{groupplot}%
	\end{tikzpicture}%
	\caption%
	{%
	\label{fig-SM_EE_PRM}%
	Entanglement entropy (black circles) of the \gls{PRM} as a function of subsystem size $l$ for different densities. %
	Blue line is a fit using the Calabrese-Cardy ansatz, \cref{CCansatz}, from which we extract the central charge $c$ given in the plots. %
 Here, no offset is used, $l_0=0$. %
	}%
\end{figure}
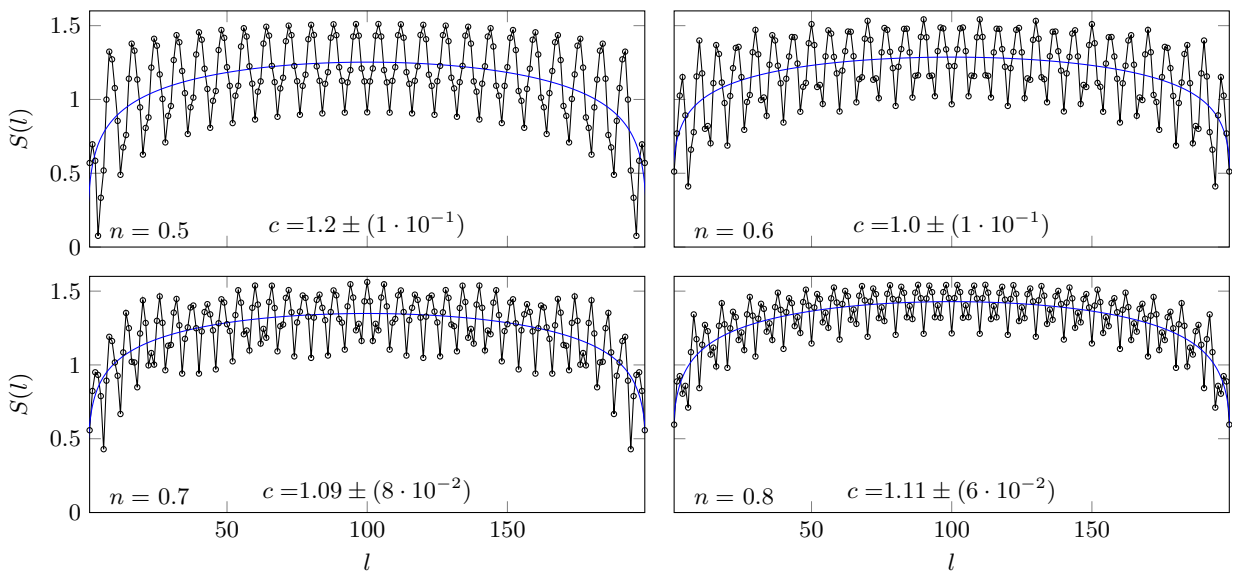%

In \cref{fig-SM_EE_PRM}, we plot the entanglement entropy as a function of subsystem size $l$ for the PRM in black circles. The entanglement entropy shows strong oscillations with several different wavelengths, which do not fit known Calabrese-Cardy formulas that incorporate oscillations, for example due to open boundary conditions~\cite{EE_osc}. A naive fit using \cref{CCansatz} is shown with solid blue line with the extracted central charge given in each subfigure.
In~\cref{fig-offset_c_vs_n_}, we show the extracted central charge $c$ as a function of density $n$ using several different offsets $l_0$. We find that the value of $c$ is often both offset-dependent and  suffers from large errors due to the strong oscillations in the entanglement entropy. Usually, whenever the Calabrese-Cardy formula works well, the effect of an offset $l_0$ should be negligible. In aggregate, while over significant segments of densities $c=1$ is not implausible, it is seemingly impossible to conclude so with any degree of certainty. There exist even some densities where $c=1$ seems highly unlikely, such as around ${ n = 2/3 }$, and for ${ n > 9/10 }$. For these reasons, we cannot identify the \gls{IP} phase of the PRM as a \gls{PLL}.

\def\offsets{0,5,10}%
\foreach \offset in \offsets%
{%
	\IfFileExists{data/SymMPS/spinlessFermions/Lx_100/EEvsDensity_offset_\offset.dat}{}%
	{%
		\immediate\write18{grep "\string^.* \offset\space" data/SymMPS/spinlessFermions/Lx_100/EEvsOffsetvsDensity.dat > data/SymMPS/spinlessFermions/Lx_100/EEvsDensity_offset_\offset.dat}%
	}%
}%
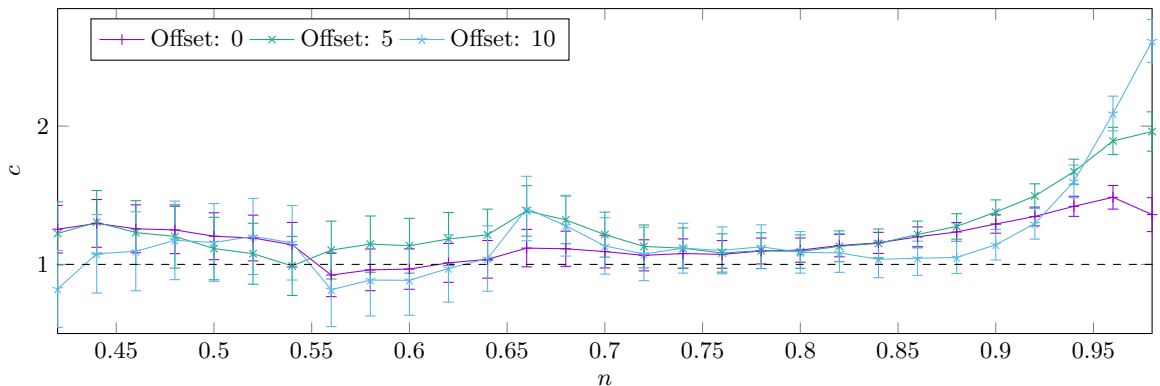
\begin{figure}[!h]%
	\tikzsetnextfilename{CompareCentralChargesVsOffsets}
	\begin{tikzpicture}%
		\begin{axis}%
		[%
			width	=	0.9\textwidth,%
			height	=	0.25\textheight,%
			legend pos = north west,%
			xlabel	=	{$n$},%
			ylabel	=	{$c$},%
			xmin	=	0.42,%
			xmax	=	0.98,%
			ymin	=	0.5,
			ymax	=	2.85,
			legend columns=3,
		]%
			\def\offsets{{0,5,10}}%
			\pgfmathtruncatemacro{\offsetsDim}{dim(\offsets)-1}%
			\pgfplotsforeachungrouped \i in {0,...,\offsetsDim}%
			{%
				\pgfmathtruncatemacro{\offset}{\offsets[\i]}%
				\pgfmathsetmacro{\marker}{\gpmarkers[mod(\i,dim(\gpmarkers))]}%
				\pgfmathsetmacro{\cl}{\gpcolors[mod(\i,dim(\gpcolors))]}%
				\edef\cmd%
				{%
					\noexpand\addplot%
					[%
						color=\cl,%
						mark = \marker,%
						error bars/.cd,%
						y dir=both,%
						y explicit,%
					]%
						table%
						[%
							x expr	=	\noexpand\thisrowno{0}/100,%
							y expr	=	\noexpand\thisrowno{2},%
							y error expr = \noexpand\thisrowno{3},%
						] {data/SymMPS/spinlessFermions/Lx_100/EEvsDensity_offset_\offset.dat};%
					\noexpand\addlegendentry{Offset: $\offset$};%
				}\cmd%
			}%
			\addplot[dashed, domain=0.4:1] {1};%
		\end{axis}%
	\end{tikzpicture}%
	\caption%
	{%
	\label{fig-offset_c_vs_n_}
	Fitted central charge for the Calabrese-Cardy formula, \cref{CCansatz}, as a function density $n$ is plotted for various offsets $l_0$.   
	}%
\end{figure}%

\section{MIM: Accuracy of analytical solution for small interaction strengths}%
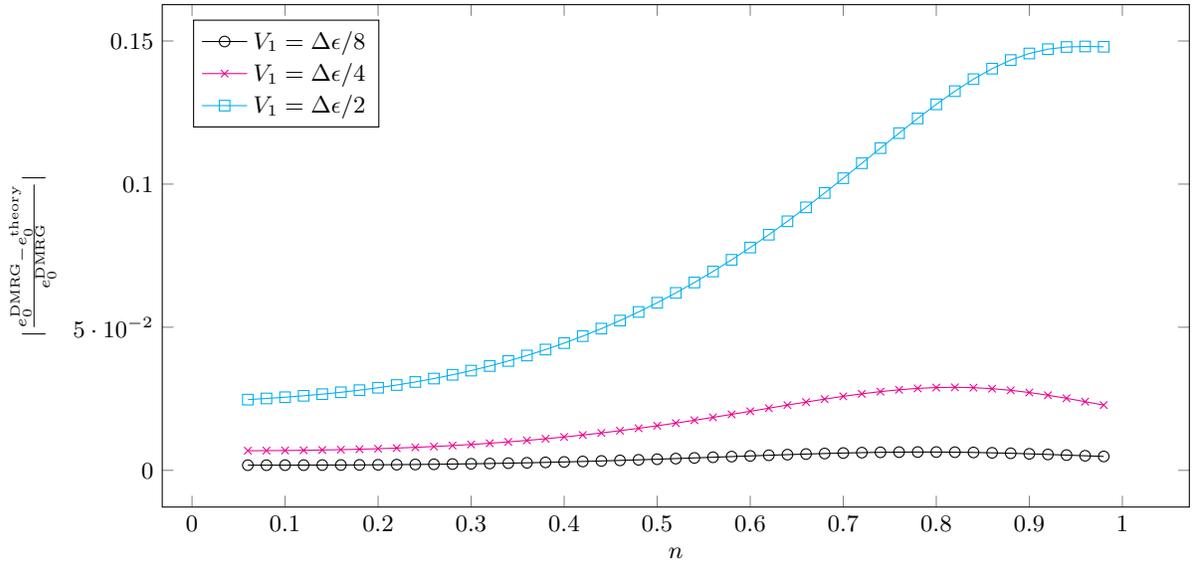
\begin{figure}[!h]%
	\centering%
	\tikzsetnextfilename{energyDifferenceVsMaciej}%
	\begin{tikzpicture}%
		\begin{axis}%
		[%
			width	=	0.85\textwidth,%
			height	=	0.35\textheight,%
			legend pos = north west,%
			xlabel	=	{$n$},%
			ylabel	=	{$\lvert \frac{e_0^{\mathrm{DMRG}} - e_0^{\mathrm{theory}}}{e_0^{\mathrm{DMRG}}} \rvert$},%
		]%
			\def\Des{{8,4,2}}%
			\pgfmathtruncatemacro{\DesDim}{dim(\Des)-1}%
			\pgfplotsforeachungrouped \i in {0,...,\DesDim}%
			{%
				\pgfmathtruncatemacro{\De}{\Des[\i]}%
				\pgfmathsetmacro{\marker}{\MIMmarkers[mod(\i,dim(\MIMmarkers))]}%
				\pgfmathsetmacro{\cl}{\MIMcolors[mod(\i,dim(\MIMcolors))]}
				\edef\cmd%
				{%
					\noexpand\addplot%
					[%
						color=\cl,%
						mark = \marker,%
					]%
						table%
						[%
							header = false,%
							x expr = \noexpand\thisrowno{0},%
							y expr = \noexpand\thisrowno{1},%
						]%
						{data/MAQUIS/SM_dmrg_vs_theory_MIM_De_\De.txt};%
					\noexpand\addlegendentry{$V_1=\Delta \epsilon/\De$};%
				}\cmd%
			}%
		\end{axis}%
	\end{tikzpicture}%
	\caption%
	{%
	\label{fig-SM_errors_MIM}
	Relative energy difference $\delta e_0$ between the quasiexact DMRG solution of the full Creutz ladder, \cref{eq-H}, and prediction by theory for the single-band limit, \cref{Eq-exact_theo_energy}, for the \gls{MIM}, as a function of density $n$ for several different interactions $V_1$.  
	}%
\end{figure}

As we state in the main text, the single-band projected Hamiltonian is exactly solvable for the \gls{MIM}~\cite{Maciej16}. In this section we compare this exact solution with the DMRG results of the full Hamiltonian. The projected Hamiltonian of the \gls{MIM} reads 
\begin{equation}
\label{Eq-exact_theo_energy}
\hat{H}_{-}=\sum_j  \epsilon_- \hat n^{\nodagger}_{-,j} + \frac{V_1}{4} \left(\hat c^{\dagger}_{-,j+1} \hat c^{\dagger}_{-,j} \hat c^{\nodagger}_{-,j} \hat c^{\nodagger}_{-,j-1} + \mathrm{H.c.} \right) \ .
\end{equation}
 The solution in Ref.~\cite{Maciej16} provides an exact analytical expression for the ground-state energy per unit cell
 \begin{equation}
 e_0^{\mathrm{theory}} = \epsilon_- n - \frac{V_1}{2 \pi} \left(1 - \frac{n}{2} \right) \sin\left(\frac{2 \pi \left(1-n \right)}{2-n} \right) \ . 
\end{equation}  
We compare this analytical expression with our DMRG results and plot in \cref{fig-SM_errors_MIM} the relative difference, $\delta e_0 = \vert \frac{e_0^{\mathrm{DMRG}} - e_0^{\mathrm{theory}}}{e_0^{\mathrm{DMRG}}} \vert$, as a function of density $n$ for several different interaction strengths $V_1$. For $V_1=\Delta \epsilon/8$, the largest $\delta e_0$ is less than $1 \%$ and for $V_1=\Delta \epsilon/4$ the maximum of $\delta e_0$ only rises to $3\% $. These small deviations justify our expectation that the single-band projected Hamiltonian, \cref{eq-projected_H} in the main text and \cref{Eq-exact_theo_energy} above, captures the physics very well. On the other hand, for larger interaction $V_1 = \Delta \epsilon/2$, the deviation from the analytical result increases monotonically with density, to as high as $15 \% $. This is not surprising as we also see a different behavior in this regime for other quantities, such as correlation functions and entanglement entropy, in the main text discussed as being due to the interplay between the two bands present in the full Creutz ladder.

\section{MIM: Phase correlation function for different interaction strengths}

\begin{figure}[!h]%
	\centering%
	\tikzsetnextfilename{CorrelationFunctions}%
	\begin{tikzpicture}%
		\begin{loglogaxis}%
		[
			width	=	0.85\textwidth,%
			height	=	0.35\textheight,%
			legend pos = south west,%
			xlabel	=	{$r$},%
			ylabel	=	{$G_{-,\mathrm{ph}}(r)$},%
		]%
			\def\Des{{8,4,2}}%
			\pgfmathtruncatemacro{\DesDim}{dim(\Des)-1}%
			\pgfplotsforeachungrouped \i in {0,...,\DesDim}%
			{%
				\pgfmathtruncatemacro{\De}{\Des[\i]}%
				\pgfmathsetmacro{\marker}{\MIMmarkers[mod(\i,dim(\MIMmarkers))]}%
				\pgfmathsetmacro{\cl}{\MIMcolors[mod(\i,dim(\MIMcolors))]}%
				\edef\cmd%
				{%
					\noexpand\addplot%
					[%
						color=\cl,%
						mark = \marker,%
					]%
						table%
						[%
							header = false,%
							x expr = \noexpand\thisrowno{0},%
							y expr = \noexpand\thisrowno{1},%
						]%
						{data/MAQUIS/SM_pair_cor_MIM_N_30_De_\De.txt};%
					\noexpand\addlegendentry{$V_1=\Delta \epsilon/\De$};%
				}\cmd%
			}%
		\end{loglogaxis}%
	\end{tikzpicture}%
	\caption%
	{%
	\label{fig-SM_MIM_Gs} 
	Superconducting phase correlation function $G_{-,\mathrm{ph}}(r)$ for the MIM at density $n=0.3$ for several different interaction strengths $V_1$. 
	Note the log-log scale. %
	}%
\end{figure}
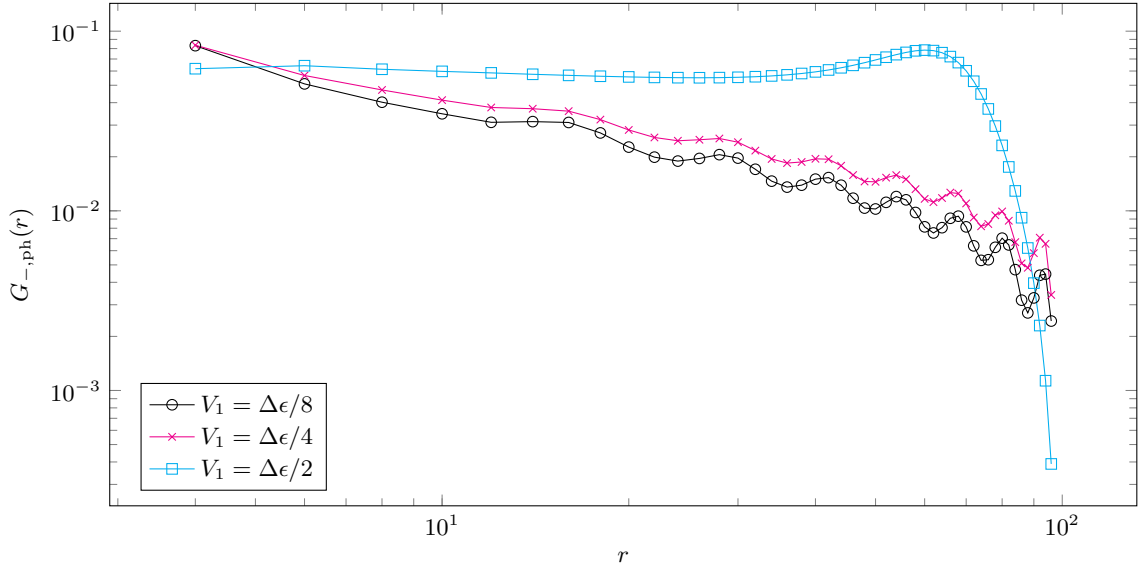

In \cref{fig-on_model_2}\subref{fig-MIM_exponents} in the main text, we present the exponent of the algebraic decay of the pair correlation function, $G_{-,\mathrm{ph}}(r)$, for the MIM at $V_1=\Delta \epsilon/2$. To show the drastically different behavior of $G_{-,\mathrm{ph}}(r)$ for this $V_1$ in comparison to smaller interaction strengths, we plot $G_{-,\mathrm{ph}}(r)$ for several different interaction strengths at fixed density $n=0.3$ in \cref{fig-SM_MIM_Gs} . It is evident that $G_{-,\mathrm{ph}}(r)$ for $V_1=\Delta \epsilon/2$ decays much slower than for lower values of $V_1$.   

\end{document}